# Spectrometer for new gravitational experiment with UCN

G.V. Kulin[1], A.I. Frank[1], S.V. Goryunov[1], D.V. Kustov[1,3], P. Geltenbort[2], M. Jentschel[2], A.N. Strepetov[4], V.A. Bushuev[5]

[1] *Joint Institute for Nuclear Research, Dubna, Russia*
[2] *Institut Lauer-Langevin, Grenoble, France*
[3] *Institute for Nuclear Research, Kiev, Ukraine*
[4] *Institute of General and Nuclear Physics, RCC «Kurchatov Institute», Moscow, Russia*
[5] *Moscow State University, Moscow, Russia*

**Abstract**

We describe an experimental installation for a new test of the weak equivalence principle for neutron. The device is a sensitive gravitational spectrometer for ultracold neutrons allowing to precisely compare the gain in kinetic energy of free falling neutrons to quanta of energy $\hbar\Omega$ transferred to the neutron via a non stationary device, i.e. a quantum modulator.

The results of first test experiments indicate a collection rate allowing measurements of the factor of equivalence $\gamma$ with a statistical uncertainty in the order of $5\times10^{-3}$ per day. A number of systematic effects were found, which partially can be easily corrected. For the elimination of others more detailed investigations and analysis are needed. Some possibilities to improve the device are also discussed.

## 1. Introduction.

Apparently, neutrons are the most suitable objects to investigate the gravity interaction of elementary particles. Although gravitational experiments with neutrons have a more than half a century history [1], the existing experimental data are quite scanty, and their accuracy is many orders of magnitude inferior to the accuracy of gravitational experiments with macroscopic bodies and atomic interferometers [2-6].

Almost fifteen years after the first observation of the neutron fall in the Earth's gravitational field [1], the gravitational acceleration was measured in a classical experiment with an accuracy of about 0.5% [7]. However, the fact of gravitational acceleration of the neutron was already earlier considered as obvious and used for precise measurements of the coherent scattering length of neutrons by nuclei. In the Maier-Leibnitz–Koester gravitational refractometer [8,9], the initially horizontal neutron beam moved parabolically, fell from height $h$ on a liquid mirror, reflected from it, and arrived at a detector. Varying the incidence height, one could determine the critical height $h_0$, at which the condition of the total neutron reflection was satisfied. In this case, $mgh_0 = V$, where

$$V = \frac{2\pi\hbar^2}{m}\rho b, \qquad (1)$$

is the effective or optical potential of the mirror, $m$ is the neutron mass, $g$ is the gravitational acceleration, $\rho$ is the volume density of atoms, and $b$ is the coherent scattering length.

Up to the mid-1970s, data on the coherent scattering lengths of neutrons on nuclei were obtained from the measurement of the neutron–atom scattering cross section, i.e., by a method that is not connected with the gravitational interaction. This allowed Koester to compare the data on the scattering length obtained by two methods and thereby to verify the fundamental principle of the equivalence of the inertial and gravitational masses of the neutron [10]. He obtained the value $\gamma = 1.0016 \pm 0.00025$ for an equivalence factor that he defined as $\gamma = (m_i/m_g)(g_n/g_0)$, where $m_i$ and $m_g$ are the inertial and gravitational neutron masses, respectively, and $g_n$ and $g_0$ are the gravitational acceleration of the neutron and the local gravitational acceleration of macroscopic bodies, respectively. More recently, Sears [11] made a number of important remarks concerning that study and repeated Koester's processing. For the equivalence factor $\gamma$, which is now defined as the ratio of the coherent scattering lengths measured by two methods, he presented a value of $1-\gamma = (3\pm3)\times10^{-4}$. Much more recently, a similar analysis was performed by Schmiedmayer [12], who obtained the equivalence factor with an accuracy twice as good as that obtained in [10]. Note that the correction of the scattering of neutrons by atomic-shell electrons was introduced in [10, 12], where data on the nuclear scattering lengths were extracted from experiments on the scattering of neutrons by atoms. This correction is in the order of 1%. At the same time, even modern data on the neutron–electron scattering length are somewhat contradictory [13, 14] and it is unobvious that their accuracy is adequate to the stated accuracy of the studies mentioned above, particularly [12] whose author noted that he used averaging of statistically incompatible data on the n-e scattering length.

The first quantum neutron gravitational experiment was performed in 1975 by Colella, Overhauser, and Werner [15], who observed the gravitationally induced phase shift of the neutron wave function in an experiment with a neutron interferometer. Results of first experiments [15, 16] basically corresponded to theoretical predictions. However, further investigations revealed certain discrepancies. In the latest work [17], the difference between the experimental and theoretical phase shifts was equal to 1% with an error smaller by an order of magnitude. The cause of this discrepancy, which remains unknown, was discussed in many theoretical studies (see, e.g., [18]). Results of a more recent experiment [19] whose accuracy was equal to 0.9% do not remove this problem. Very recently the of the experiment with a neutron spin-echo spectrometer was performed [20]. Author reported that their experimental result for the gravitation induced phase shift agrees within approximately 0.1% with the theoretically expected result, while the overall measurement accuracy is 0.25%.

Another quantum gravitational experiment with neutrons was performed recently [21, 22], while the possibility of this experiment was predicted earlier in [23]. Nesvizhevsky et al. [21, 22] reported the observation of the quantization of the vertical-motion energy of ultracold neutrons (UCNs) stored on a horizontal mirror. It is possible to hope that detailed investigation of this effect or a similar phenomenon accompanying the storage of UCNs over the magnetic mirror [24] will be very useful for studying the gravitational interaction of the neutron as a quantum particle.

Very promising results were recently obtained by T. Jenke et al. [25]. They observed transitions between quantum states of UCN being stored on a plane mirror. An original approach to the test the equivalence principle for neutrons was proposed recently in [26]. The idea was based on the huge sensitivity of the neutron-diffraction method to the force acting on a neutron.

In the experiment [27] the change in the energy of a neutron falling to a known height in the Earth's gravitational field was compensated by a quantum of energy $\hbar\Omega$ that was transferred to the neutron via a non-stationary interaction with moving diffraction grating. The aim of the experiment was the determination of the value $\gamma = mg/(mg)_{exp}$ where $m$ is a table value of the neutron mass, $g$ is the local free fall acceleration and $(mg)_{exp}$ the gravity force acting on a neutron found in the experiment. In the experiment the value of the equivalence factor was found to be $1-\gamma = (1.8 \pm 2.1)\times 10^{-3}$.

In the present paper we describe the experimental set up for a new experimental test of the equivalence principle for neutrons. Results of the test experiment and some future prospects of improvements of the device are also discussed.

## 2. Moving diffraction grating as a nonstationary device for the neutron energy transformation.

As in [27] in the new experiment the change in energy $mgH$ of a neutron falling from a known height in the Earth's gravitational field is compared to a known quantum of energy $\hbar\Omega$, which is transferred to the neutron by a non-stationary device. As latter a moving diffraction grating playing the role of a quantum phase modulator is used. The phenomenon of energy quantization via diffraction of the neutron by a moving grating was first predicted in [28] and experimentally observed in [29-31].

We present here briefly the theoretical description of the correspondent quantum mechanical problem. In the laboratory system of reference a plane neutron wave $\Psi_{in}(x,z,t) = A_0 \exp(ik_0 z - i\omega_0 t)$ is assumed to propagate along the z-axis towards a periodic grating, which assumed to be normal to the neutrons propagation axis (fig.1a). We denote with $k_0 = m v_0/\hbar$ the neutron wave number, where $m$ is the neutron mass, $\hbar$ the Planck constant and $\omega_0 = \hbar k_0^2/2m$. The grating is moving with constant speed $V$ along positive direction of the $X$ axis (fig.1b) and grating grooves are oriented along $Y$ axis. Following ref. [28] we shall solve the problem in the moving frame of reference $(x', z)$ where the grating is in rest. In this system the wave incident the grating obliquely:

$$\psi'_{in}(x',z,t) = A_{in}(x') \exp[i(k_0 z - \omega' t)], \qquad (2)$$

with $A_{in}(x') = A_0 \exp(-ik_V x')$, $k_V = mV/\hbar$, $\omega' = \omega_0 + \omega_V$, $\omega_V = \hbar k_V^2/2m$. After passage of the grating in the region $z > 0$ we have

$$\Psi'(x',z,t) = A_0 \int_{-\infty}^{\infty} F(q) \exp(iqx' + ik_z z - i\omega' t) dq, \qquad (3)$$

where $k_z = (k_0^2 + k_V^2 - q^2)^{1/2}$. $F(q)$ is the Fourier transform of the product $F(x) = A_{in}(x)f(x)$ and $f(x)$ is the grating transmission function. For a periodical function $f(x) = \Sigma_j a_j \exp(iq_j x)$ where $j$ are integer numbers, $q_j = 2\pi j/L$, and $L$ is the space period of the grating, we obtain for the Fourier transform $F(q) = A_0 \sum_j \delta(q - q_j + k_v)$, and the Fourier coefficients are

$$a_j = \frac{1}{L} \int_0^L f(x) \exp(-iq_j x) dx, \quad j = 0, \pm 1, \pm 2, \ldots \qquad (4)$$

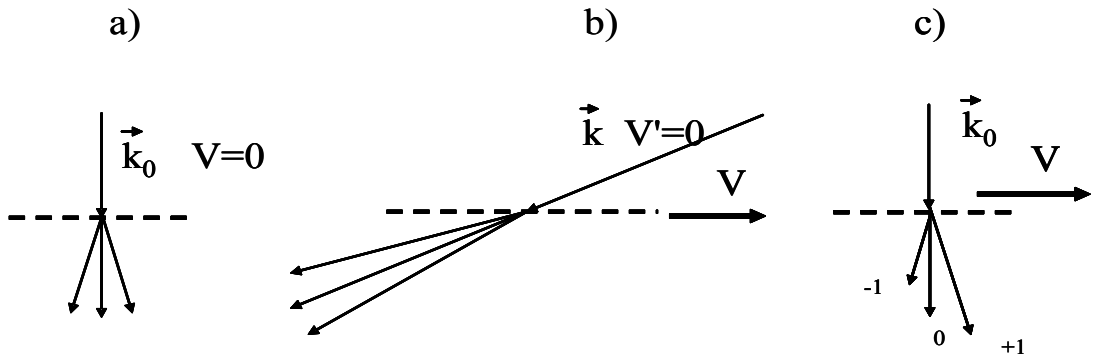

Fig. 1 Diffraction by a grating. $a$ – grating in rest, $\omega_j = \omega_0$; $b$ – reference frame is moving together with grating, $\omega_j = \omega'$; $c$ – moving grating in laboratory frame of reference, $\omega_j = \omega_0 + j\Omega$.

Passing back to the laboratory frame of reference $x = x' + Vt$ we obtain a superposition of plane waves with amplitudes $b_j$, discrete frequencies $\omega_j$ and wave vectors $\mathbf{k}_j = (q_j, k_{zj})$ (see fig. 1c):

$$\Psi(x,z,t) = \sum_j b_j \exp(iq_j x + ik_{zj} z - i\omega_j t), \qquad (5)$$

where $b_j = a_j \left[ (k_0)^{1/2} / (k_0^2 + 2k_V q_j)^{1/4} \right]$, $k_{zj} = (k_0^2 + 2k_V q_j - q_j^2)^{1/2}$, $\omega_j = \omega_0 + j\Omega$, $\Omega = 2\pi/T$, $T = L/V$.

Let us consider a grating with a rectangular groove profile with width $L/2$ and depth $d$, such that the phase difference $\Delta\varphi = k_0(1 - n)d = \pi$, where $n$ is the refraction index for neutrons. In the case of normal fall the amplitudes $a_j$ are

$$a_j = \begin{cases} 0 & \text{at } j = 2s \\ i\left(\dfrac{2}{\pi j}\right) & \text{at } j = 2s - 1 \end{cases}. \qquad (6)$$

It is obvious that the amplitudes for the even diffraction orders, including the zero, vanish and that the major part of the flux concentrates in $\pm 1$ orders.

Obviously, for larger values of the spectral splitting $\Omega = 2\pi V/L$ it is necessary to increase the grating speed $V$ and (or) decrease the space period $L$. But due to the oblique incidence in the moving reference frame the phase profile varies with increasing velocity and takes the form of trapezium [32] (see fig.2).

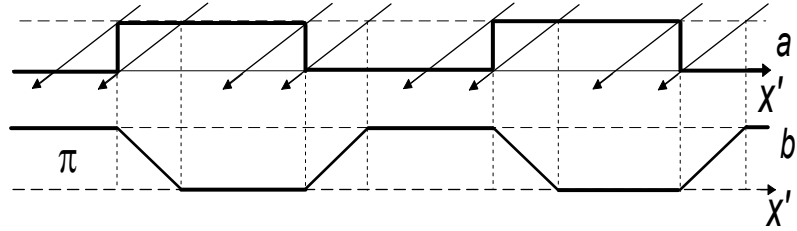

Fig. 2. $a$ – profile of grating, $b$ – profile of phase.

From (4) it follows that amplitudes $a_j$ are now

$$a_j = \begin{cases} cB_j & \text{at } j = 2s \\ -B_j/j & \text{at } j = 2s - 1 \end{cases} \qquad (7)$$

where

$$B_j = [1 + \exp(-i\pi j c)]/[i\pi(1 - j^2 c^2)], \quad c = \dfrac{2dV}{Lv_0}. \qquad (8)$$

The dimensionless parameter $c$, introduced in [32], defines the degree of the phase profile distortion. At $c = 1$ the phase profile takes the form of a triangle. Strictly speaking in the case of the moving grating the amplitudes of the zero and all even orders never vanish and their intensities increase with increasing of c, contrary to intensities of the odd orders which decrease.

## 3. Neutron Fabry – Perot interferometers and other multilayered structures for UCN spectrometry

Neutron Fabry-Perot interferometers (FPI) also referred as neutron interference filters (NIF) as well as more complex multilayered structures were many times used for the UCN spectrometry [29-40]. In context with the device to be discussed in the present paper they also play an important role. In the simplest case NIF represent a three layered structure of thin films coated on a substrate, which is transparent to UCN. The materials for the filter are chosen such that the effective potential of neutron–matter interaction for the outer layers is larger than the potential of the inner layer. Consequently the potential structure of the filter has a two-humped barrier with a well in between. Assuming the thickness of the inner layer not to be too small compared to the normal component of the neutron wave length the well width is enough for the formation of quasi-bound states. In the vicinity of these levels the structure becomes transparent for neutrons with energies below the barrier. In this case the transmission function of this structure has a sufficiently resonant character. The width of the resonance is determined by the penetrability of the outer barriers and its position may be obtained by matching the wave functions at the boundaries. Note that all the characteristics of the resonance depend only on the normal component of the neutron wave-vector.

A. Seregin first proposed in 1977 to prepare a three-layer structure with a potential, which has a form of two humped barriers [41]. In 1980 the resonant tunneling of the UCN through the three-layer filter was demonstrated experimentally by A. Steyerl et al. [42] and later by M. Novopoltsev and Yu. Pokotolovsky [33].

As it is known from quantum mechanics, a system of two coupled resonators is usually characterized by the splitting of the levels. This phenomenon was also demonstrated in an experiment [42, 43, 34]. Obviously this is the case of five-layer filters for which the potential has the form of three barriers with two wells. It is worth to mention that the value of splitting is defined by the degree of the resonator coupling, in the present case by the permeability of the central barrier. Five-layers structure can be designed such that the value of splitting would be of the same order as the natural width of the levels. In this case the system will have effectively only one transmission line. At the same time the increase of the total thickness of the barriers when compared to tree layers structures yields a remarkable suppression of the transmission curve wings.

The following main kinds of the NIF were used in our spectrometer:

1. Five-layers NIF – acting as monochromator with sole transmission line, having theoretical width of about 3.5 neV. The calculated curve of its transmission over a relatively wide wave length scale is graphed in figure 3.

2. Nine-layered filter – acting as window with relatively wide (10-12 neV) transmission band intended for the extraction of the minus first order diffraction line from the neutron spectrum raised at UCN diffraction by a moving grating. The calculated transmission curve of one of them together with the calculated transmission curve of the monochromator is shown in figure 4.

3. A special filter, named "superwindow" [34, 35], which transmits UCNs with energies in the range of 50-200 neV and effectively reflects neutrons with energies in the range of 200-700 neV. It is based on a M=2 supermirror with some added antireflection layers and is used for the suppression of the background of neutrons with energy higher than the boundary energy of nickel.

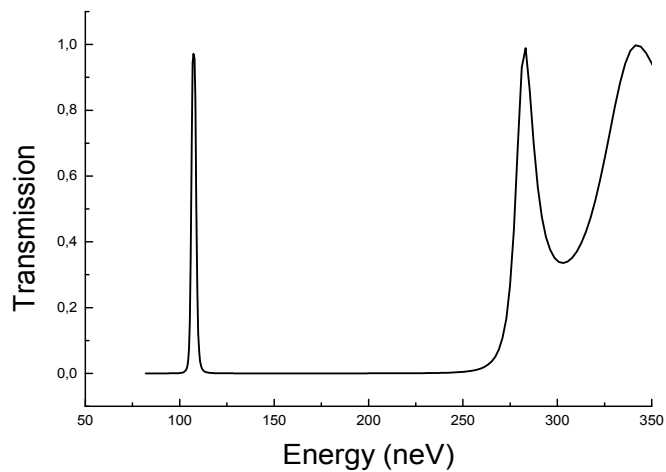

**Fig.3**. Calculated transmittivity of the 5-layer filter as a function of the "normal" energy $E_\perp = \left(\hbar^2/2m\right)k_z^2$ where $k_z$ is the component of the wave vector, normal to the filter surface

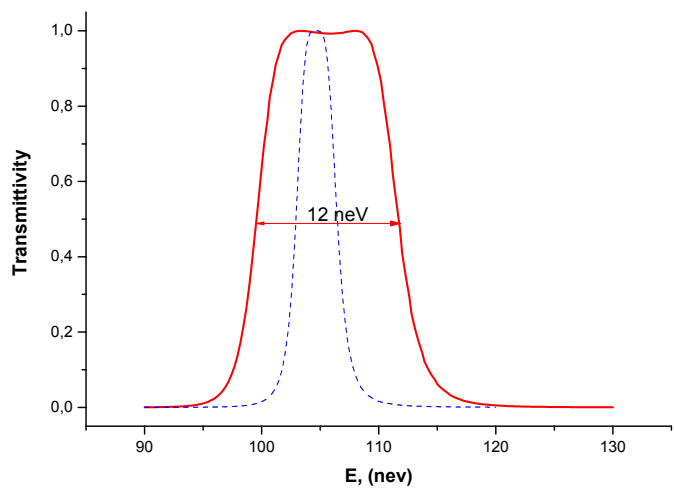

**Fig.4**. Calculated transmittivity of the 9-layer with wide transmission band together with the calculated transmittivity of the 5-layers filter.

As a material with high and zero coherent scattering amplitudes respectively we used the non-magnetic alloy Ni(0.8)Mo(0.2) and a Ti/Zr alloy. All structures were manufactured by magnetron sputtering on Si wafers of 150mm in diameter and 0.6 or 0.3mm in thickness.

## 4. The spectrometer and procedure of the gravitational experiment

The experimental installation for the gravitational experiment is shown in figure 5. As in the previous work [27], neutron interference filters are used as a monochromator and an analyzer. A controllable decrease in the neutron energy is achieved by a diffraction on a moving grating. In contrast to [27] the energy of neutrons is measured now by a peculiar time-of-flight method as was proposed in [44]. For this purpose the neutron flux is modulated by the chopper and the detector measures the corresponding oscillation of the count rate. The phase of the count rate oscillation $\Phi = 2\pi\chi\tau$ is defined by the frequency of chopping $\chi$ and the time of flight $\tau$. The main benefit of this approach is that the modulation phase does not depend on UCN beam intensity and therefore is not sensitive to fluctuations of the background or the detector efficiency.

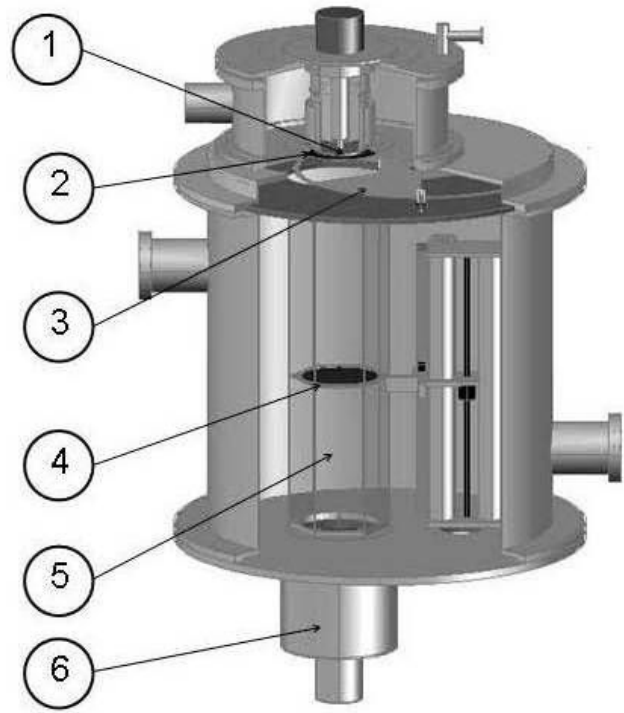

**Fig.5.** Experimental installation. 1 - Fabry-Perrot monochromator, 2 – diffraction grating, 3 - rotor of the chopper-modulator, 4- movable FPI-analyzer, 5 – vertical glass neutron guide, 6 – scintillation detector

As before we use a rotating grating prepared on the surface of a silicon disc 150 mm in diameter and 0.6 mm in thickness. Radial grooves were prepared in the peripheral region of the disc, which is a ring with an average diameter of 12 cm and a width of about 2 cm. The widths of the grooves are proportional to the radius and this proportionality ensures a constant angular distance between the grooves equal to a half period. The angular period of the structure is exactly known to be $\alpha = 2\pi/N$ with $N = 75\,398$. The depth of the grooves is equal to 0.14 μm and it was chosen to ensure a phase difference $\Delta\varphi = \pi$ between the neutron waves passed through neighbor elements of grating as it was explained in section 2. The grating was manufactured by the Qudos Technology Ltd.

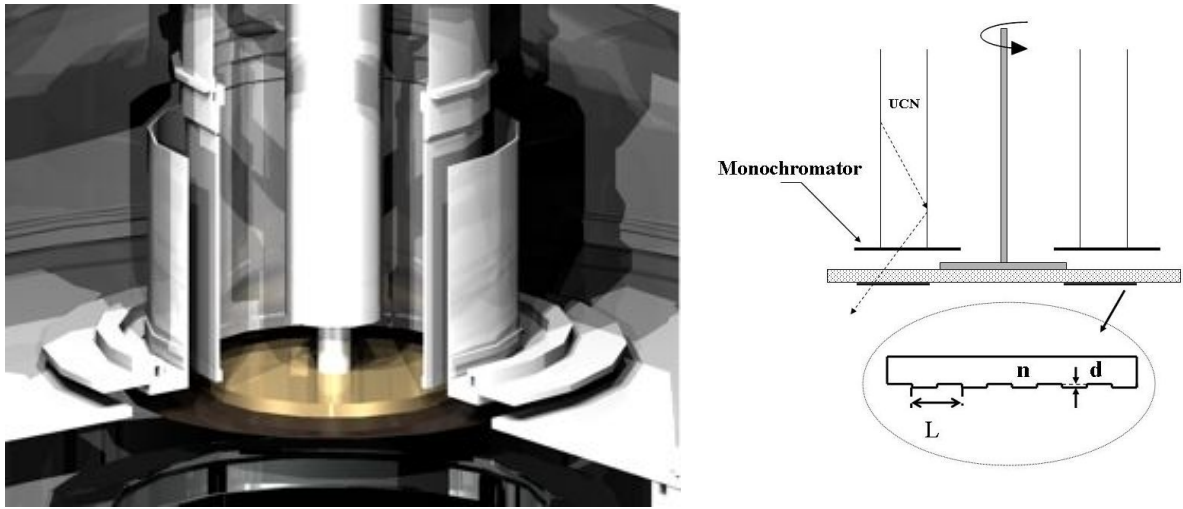

**Fig. 6.** Upper part of the spectrometer. The annular corridor, providing UCN to the spectrometer, IFP-monochromator and grating are shown

Below the upper part of the spectrometer the chopper acting as modulator of the neutron flux (3 in fig.5) is placed. The first version of the modulator was fabricated in 2010 and tested at the UCN source of Institute Laue-Langevin [45]. In these first experiments the perspective of a modulation approach to time-of-flight UCN spectrometry was demonstrated [46]. However, it was found that modulator itself needed an improvement, which was done one year later.

The main element of the new version of the modulator is a rotor made from a titanium disk of 2 mm in thickness. It has three open sectors (see fig.7a) and when being rotated it opens and closes periodically the way of UCNs towards the vertical glass neutron guide with entrance hole of 150 mm in diameter. The rotor is driven by a Phytron stepper VSS-65 HV located outside of the vacuum volume coupled to the rotor via a tooth belt and magnetic coupling. The rotation frequency of the rotor may reach 1800 rpm, which corresponds to a modulation frequency of 90 Hz. For the use of the electronic control system an infrared

Honeywell sensor VSS-65 HV and a small slot at the periphery of rotor are used. Stability of the rotor rotation frequency is of the order of $10^{-5}$.

UCNs, passing the modulator are coming to the vertical neutron guide, formed by the six 95x680мм flout glass plates (pos.5 in fig.5), and reach the scintillation detector. A carriage with an analyzing filter may be placed inside the neutron guide. The position of the carriage can be changed in height via the use of a ball screw. The screw shaft can be rotated by a stepper motor, which is placed outside the vacuum vessel and coupled with it via a magnetic coupling.

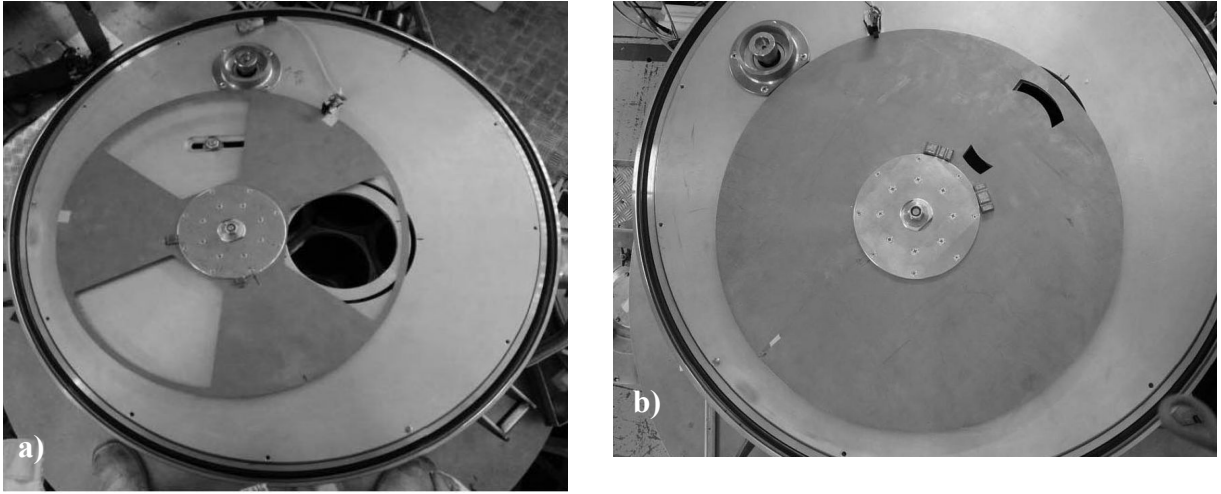

**Fig. 7**. Rotors of the chopper. a) Rotor for the periodical modulation. b) Rotor for usual TOF measurements

The experimental procedure may be divided into two independent stages. The first step is the phase calibration. At this stage the FPI is placed in a movable carriage and the phase of count the rate oscillation (PCRO) $\Phi_{an}(H)$ is measured as a function of the analyzer position $H$. At the second step, which is the main part of the experiment, the NIF-monochromator is placed in the upper part of the spectrometer as shown in fig 6. Before entering the vertical neutron guide the UCNs pass through the rotating phase grating. It provides a high-frequency neutron wave modulation resulting in the appearance of sidebands in the neutron energy spectrum. In order to transmit only neutrons with -1 diffraction order to the detector (those shifted in energy by $\Delta E = -\hbar\Omega$), the 9-layers analyzer filter with relatively wide transmission band is placed into the movable carriage. Here we use $\Omega = 2\pi f N$, where $f$ is the frequency of grating rotation. The circular frequency of the wave modulation may rich $4.7\times10^7$ rad/s, which corresponds to an energy transfer of $\Delta E \approx 30$ neV. During this phase of the experiment the dependence of the PCRO $\Phi_{mon}(\Omega)$ is measured in the following way: For a number of measurements with different frequencies $\Omega_i$ it is possible to determine those height positions of the 9-layers analyzer $H_i$, which satisfy the equation $\Phi_{mon}(\Omega_i) = \Phi_{an}(H_i)$, and finally obtain a set of equation $\Delta H_i = \hbar\Omega/mg_n$, where $mg_n$ is the gravity force acting at the neutron.

## 5. Results of test experiments.

### 5.1. Testing of the chopper-modulator and spectrometric properties of the device.

The second version of the chopper-modulator was tested at the same UCN source as the first one. In figure 8 the oscillation of the detector count rate measured at modulation frequency 90Hz [47] is shown. Five layers NIF with a theoretical value for the transmission line of 107 neV and a FWHM of the order of 4 neV was used as a monochromator. Due to a high degree of monochromatization the detected amplitude of the count rate modulation was relatively large, even when the time of flight was ten times greater than period of modulation.

In figure 9 the fitted value of the PCRO in dependence on modulation frequency is shown. Measurements were done when the monochromator was placed 20 cm below the annular corridor. The result demonstrates the high degree of linearity and the total time of flight may be easily found from the line slope. Notice that the relative accuracy of the measured time of flight was $5.5\times10^{-4}$, while total measurement time was less than 15 min.

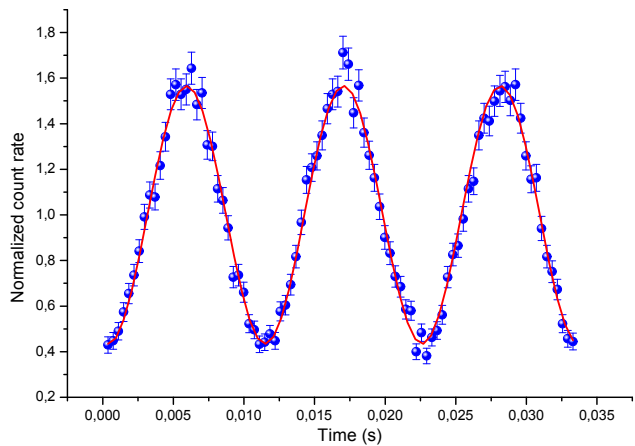 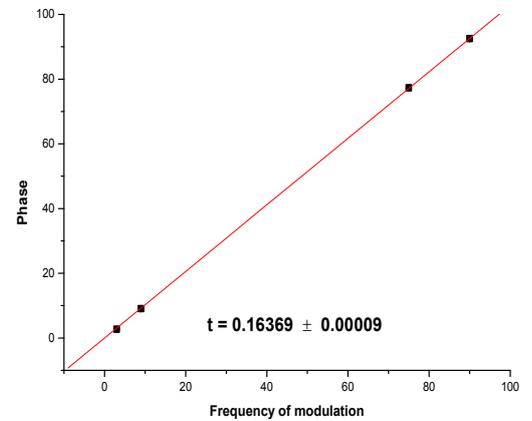

**Fig. 8.** Oscillation of the count rate fitted by sine function. Modulation frequency 90 Hz.

**Fig.9**. Phase of the count rate oscillation in dependence on the modulation frequency

For the using of spectrometer in usual time-of-flight mode another rotor blade for the chopper was made. It has two arcform slots bordered by the sector with an angular dimension of $2\pi/18$ (see fig.7b). When it was in use the entrance aperture of the glass neutron guide was masked by stator diaphragm with an open sector having the same angular dimension. For quantitative estimation of the spectrometric properties of the device the time of flight spectrum formed by NIF was measured. The monochromator was placed at its standard position, at the exit of the annular corridor, and the grating and analyzer carriage were removed. In figure 10 the results of such measurement are presented. The FWHM of the time of flight peak was found to be 3.7ms, which corresponds to a relative time spread of $\Delta t/t \approx 0.027$. Taking into account that the FWHM of the chopper operation function was about 2.4 ms and total time of flight was 139 ms for the width of the monochromator spectrum we obtain $\Delta v/v \approx 0.02$, which is in a good agreement with the calculation.

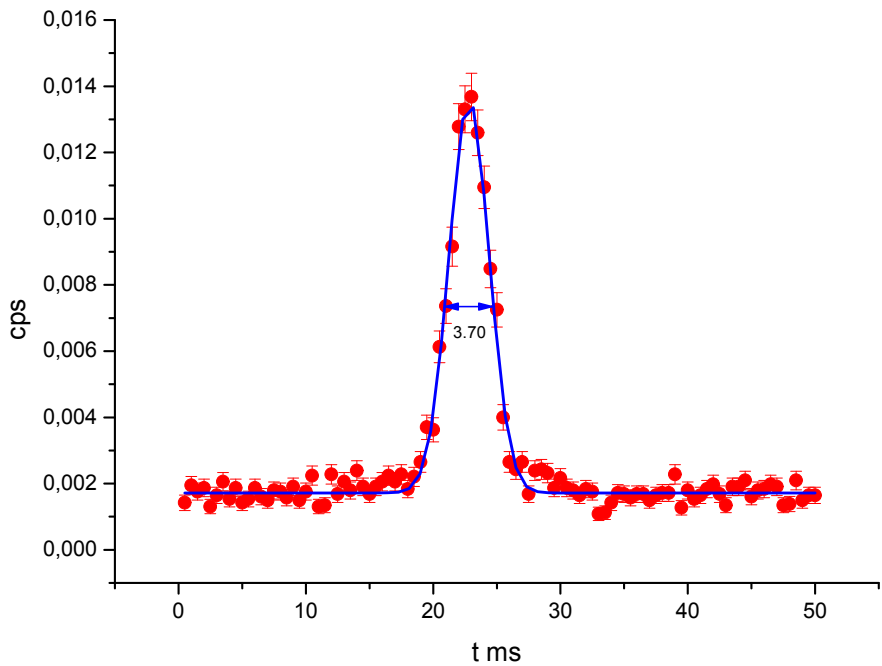

**Fig. 10.** Time of flight spectrum of neutrons, passed through FPI-monochromator plotted at the one period of chopper rotating**.**

## 5.2. Testing of the experimental procedure

First experiments to test both stages of the planned gravitational experiment were performed in 2011 [47]. For the test of the calibration stage the NIF – monochromator was placed in the carriage (pos.4 in fig.5) and the PCRO was measured as a function of the carriage position in height. For the suppression of background caused by the very cold neutrons, an additional superwindow filter was combined with the monochromator. Results of this measurement are shown in figure 11. Note that the error bars are smaller than the size of the points. These latter were typically 0.03-0.05 radian, while the total phase was more than 70 radian.

The second, main stage of the experiment was also tested. The NIF- monochromator was installed at the exit of the annular corridor just above the rotating grating (pos.2 in fig.5). Two filters were placed at the carriage: the nine layers window-filter and the 110-layers superwindow. This pair served as analyzer.

When the analyzer was placed in a proper height position it transmitted UCNs, which passed through the monochromator changed their energy due to -1 order diffraction at the moving grating, and finally accelerated by gravity. The PCRO as a function of carriage position might be measured. The result of such measurement corresponding to the gratings rotation frequency of 6300 rpm is shown in fig. 11 together with results of the calibration stage. As it was explained above, the aim of such a measurement was to find the crossing point of two obtained lines.

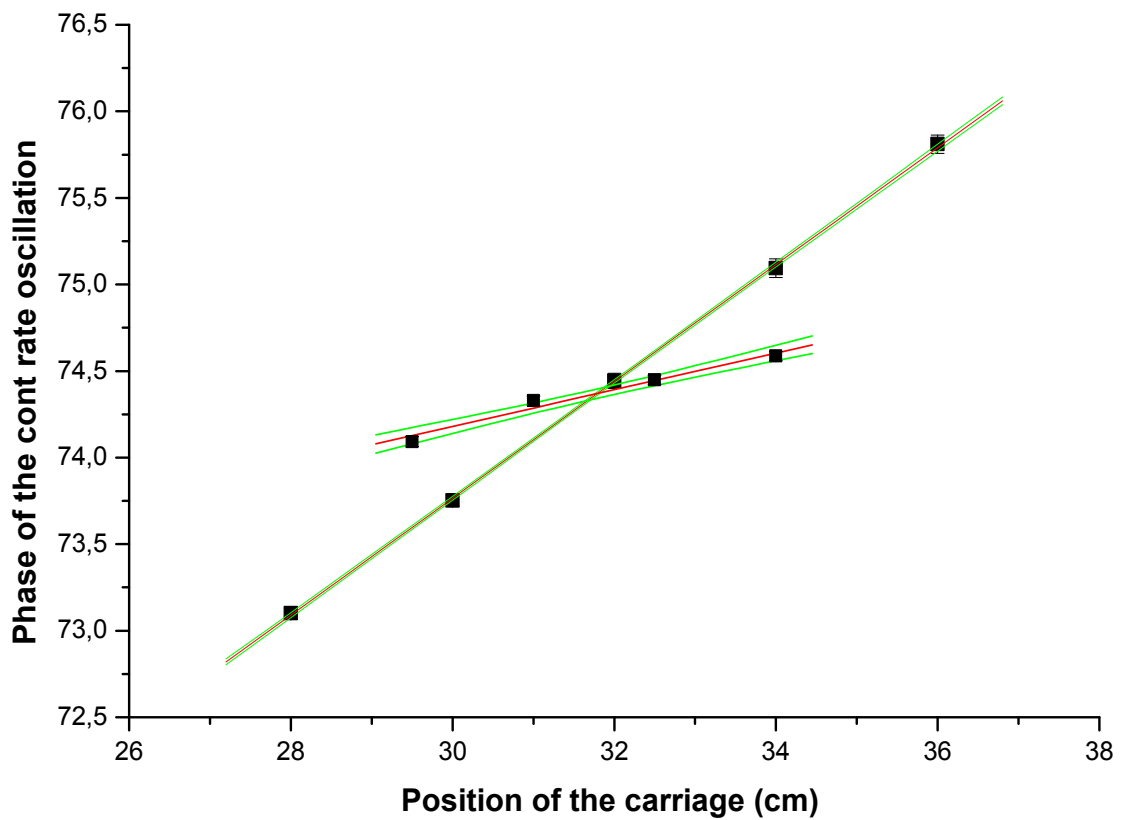

**Fig. 11**. Phases of the count rate oscillation measured in two stages of the experiment [46]. Modulation frequency 75Hz.

A surprising result was that the PCRO depended much stronger on the position of the analyzer than expected and that the angle between the two phase curves was relatively small. Obviously, the position of a crossing a point between two lines can be more precisely determined, when of the angle between the lines increases. Thus the main problem of the beam-time of 2011, can be identified to be the very the low accuracy collection rate for the equivalence factor, which can be estimated to be in the order of $1.5 \times 10^{-2}$ per day.

To improve the experimental conditions several modifications were done before the next beam time in 2012 [48]. To decrease the total thickness of material on the way of UCNs we used new multilayer structures deposed at 0.3mm Si wafers, which was half the thickness of what was used before. Some steps were taken to suppress the background caused by neutrons which might bypass the analyzer on the way to detector. As a result the statistical conditions of the measurements became much better than in previous experimental cycle (see figures 12).

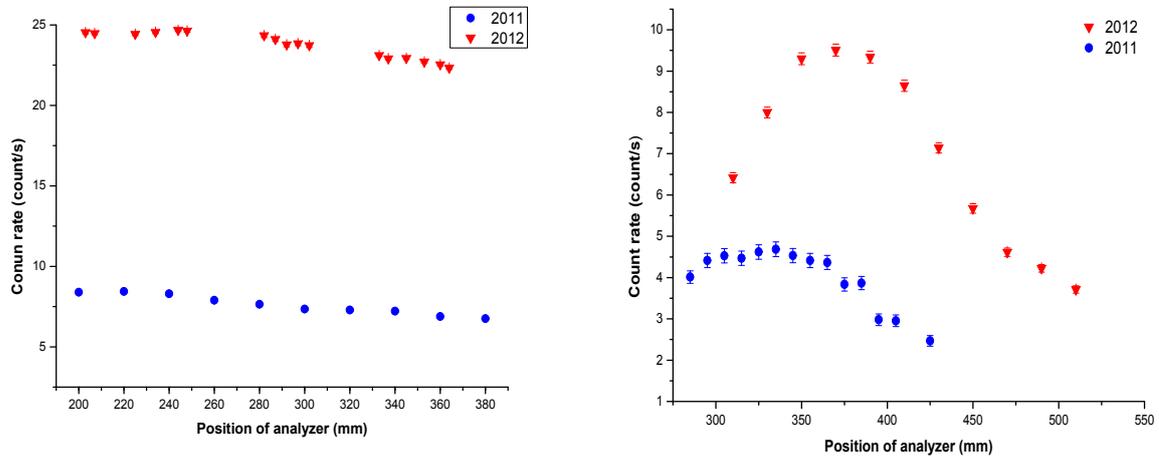

**Fig.12.** Comparison of experimental conditions in 2011 (dots) and 2012 (triangles). In the left: count rate in the geometry of calibration. In the right: scanning curve of - 1 diffraction order made using 9-layers filter. Grating rotation frequency was 6300 rpm.

Both stages of the experiment were tested again. The use of a new monochromator, characterized by a narrower transmission spectrum, permitted us to increase the operation frequency of the chopper-modulator up to 84Hz without decreasing the modulation amplitude.

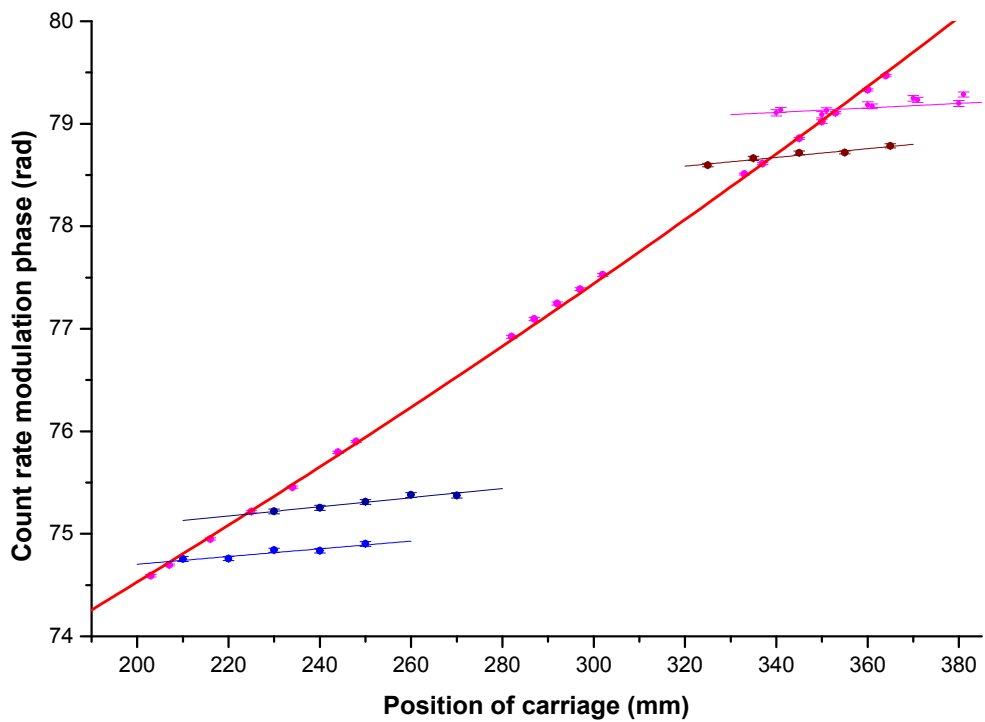

Fig.13. Phases of the count rate oscillation measured in two stages of the experiment [47]. Four fitted lines correspond to the measured PCRO in second (main) stage of the

The main stage of the experiment was performed with the grating rotating at 3600, 3900, 6000 and 6300 rpm. The width of the 9-layer analyzer transmission band was increased up to 12 neV. That increased the count rate in the main stage of the experiment and, what is more important, increased an angle between the two crossing lines of the phases, which were obtained in the two stages of the experiment.

The results of both measurements are displayed in figure 13. The estimation of the statistical accuracy collection rate under these new conditions is $5 \times 10^{-3}$ per day. This rate is enough to collect a statistical accuracy in the order of $5 \times 10^{-4}$ within 80 days (two reactor cycles). This is the main positive result of the experiment, while at the same time a number of problems were identified. They will be discussed in the next section

### 5.3. Identified problems and systematic effects

As can be seen clearly in fig.13 the two fitting lines, corresponding to rotation frequencies 3600 and 3900 rpm, cross the calibration curve outside the actual measurement range. This happened since theoretically predicted positions of the crossing points did not match the experimental crossing points.

The nature origin of this systematic effect is probably due an admixture of the zero diffraction order to the peak of the minus first order that should be separated by a wide-band analyzer. It increased the averaged energy of transmitted neutrons and therefore decreased time of flight and phase of their count rate oscillation. That corresponds to a downward displacement of the experimental points in fig.13.

In conditions of a relatively wide transmission band of the analyzer this admixture of the zero order could occur due to various reasons. One of them is trivial and was identified only after the experiment. Due to an unfortunate mistake the inner diameter of the ring grating structure was exactly equal to the inner diameter of annular corridor which transports UCN to the grating (see fig. 5). Consequently some fraction of UCNs could enter into the vertical neutron guide around bypassing the grating structure.

Another circumstance is that in conditions of our experiment the parameter $c$ in (8) is not small enough and the amplitudes of even orders may differ from zero. At present we do not have any reliable data concerning the transmission spectrum of the used nine layered analyzer. All that gives freedom for various speculations and only future investigations may shed light on the problem.

Another problem revealed itself after analyzing the time of flight dependence on the NIF position obtained in the calibration measurement. The time of flight value may be calculated according to geometrical dimensions of the spectrometer and the energy of UCNs as transmitted by the monochromator. This latter was defined by a) an absolute measurement of the time of flight spectrum of neutrons passing through a "standard" monochromator and by b) comparison of the transmission spectrum of the "standard" with the spectrum of UCNs passing through a monochromator under investigation. With this aim two scanning curves of both monochromators with the same analyzer were measured. Here it is worth mentioning that the measurement of the scanning curve is a is a standard procedure of gravity UCN spectrometers [36].

Calculated TOF curves obtained with these data differed significantly from the experimental results (see fig.14). One reason for such discrepancy is the finite thickness of the chopper rotor, which was 2 mm. Moving across the UCN flux the rotor absorbs those neutrons, which collide with it. Therefore the effective opening time was decreased at the time $\Delta t = v/d$ where $v$ is the neutron velocity and $d$ is the rotor thickness. When the corresponding correction was made, the calibration curve shifted down without changing its form. Still, the disagreement with the experimental point is significant.

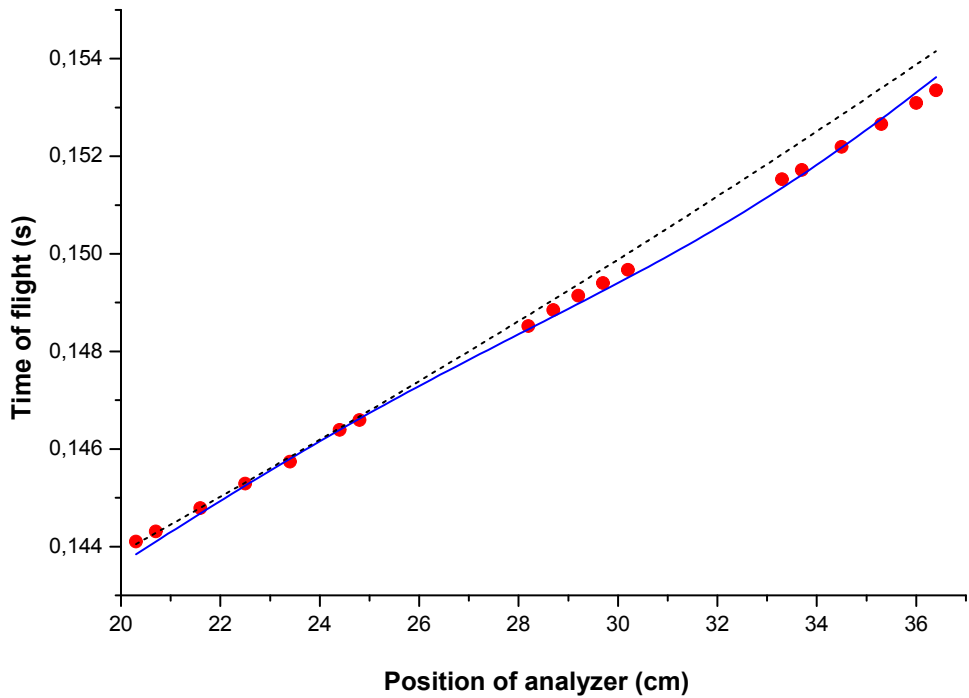

**Fig.14**. Dependence of time of flight on the position of analyzer. Dots – experimental results shown in fig. 13 but recalculated into the time scale. Dotted line – calculated curve before corrections. Solid line – corrected calculated curve (see text)

Apparently some other sources of systematic effect exist. One hypothesis was that the chopper might slightly modulate the background of those thermal or cold neutrons, which are produced by inelastic scattering of UCNs at any elements of the spectrometer. Due to the large velocity of such neutrons they reach detector within relatively short times and result in a count rate oscillation with a constant phase. In this case the count rate of the detector is the sum of two oscillation fluxes: UCN flux with a phase depending on the analyzer position and background neutron flux oscillating with constant phase. Calculations show that for a satisfactory matching with experimental data it is enough to assume the presence of an oscillating background with zero phases and amplitudes in order of 0.2 per cent of the useful signal amplitude. The resulting calculated curve, which takes both corrections into account, is shown in fig.14

## 6. Possibilities for the improvement the spectrometer

### 6.1. The problem of even orders admixture to the spectrum of minus first order

Probably the main problem, which must be solved for the success of a future gravitational experiment, is the problem of a clean extraction of the minus first order from the spectrum of diffracted neutrons. Unfortunately it was underestimated during the initial conception of the experiment and requires serious efforts in various fields to be made.

a) It is necessary to exclude firstly the possibility for UCN to reach the neutron guide and detector while bypassing the grating. As a consequence the area of the grating structure will be increased.

b) As it was noted above, the intensities of zero and second order do probably differ from zero (strictly speaking they are unknown). Unfortunately, the phenomenon of neutron diffraction by a moving grating is not investigated good enough. The theory presented in section 2 is very approximate and in the experiment [31, 32] only intensities of ± 1 orders were observed. In this experiment the grating period was four time greater than now and therefore the factor $c$ was four time smaller. We are presently working to change this situation. More rigorous theoretical calculations are in progress and an dedicated experiment aiming to measure the UCN spectrum after diffraction from a moving grating is scheduled in the near future. In the case of a successful experiment important information concerning the intensities and widths of the spectra of some lower orders will be obtained for various values of the parameter $c$.

c) For the extraction of the minus first order we initially planned to use filter-window with a calculated spectral transmission width of 8-10 neV. For technical reasons we have been later forced to use a filter with a greater width. By this we increased the risk of an admixture of the wrong neighboring orders. But the detected weak linear dependence of PCRO on the position of this analyzer (see fig.13) testifies that the first order spectrum is still not completely overlapping with the transmission spectrum of analyzer. Consequently either the form of the first order spectrum or of the transmission function of the wide band filter differs from the calculated values. Consequently along with investigation of the diffraction spectrum we plan also to specifically investigate the properties of the multilayered analyzer.

d) The problem may be solved completely if the splitting of lines in the diffraction spectrum could be increased sufficiently, for example by two or three times. The only way to do this is to increase the modulation frequency of the neutron wave. It is defined by the ratio of the diffraction gratings velocity $v$ and period $L$. Leaving aside the problem of a substantial increase the production price of the grating, we have to notice that phase grating of the currently used type will have a rather poor efficiency. The reason is the dramatic increase of the parameter $c$, which is proportional to the same ratio $v/L$. Nevertheless this possibility needs to be analyzed in more detailed. This will only become possible after completing the planned program of investigating the UCN diffraction by a moving grating.

## 6.2. On the possibility of increasing sensitivity of the planned gravitational experiment

Of course it would be very useful to increase the sensitivity of the experiment. From this point of view it is preferable to increase the range of gravitational energy $mgH$. For this it is necessary to proportionally increase the range of energy, which is transferred to neutrons by a moving grating. Unfortunately this possibility is limited. Decreasing the frequency of modulation and spectrum splitting yields an increased risk of the zero order admixture, while increasing the frequency decreases the diffraction efficiency of the grating. From this point of view the search of another type of grating with small profile height becomes very important.

Another way is to increase the slope of the calibration curve. With this aim we plan to increase the height of the vertical neutron guide and simultaneously decrease the initial energy of the UCNs to approximately 95 neV, which requires a new monochromator.

Probably we do not yet exhaust all possibilities to increase the count rate of UCN. Different ways to increase the spectrometer luminosity are also currently analyzed.

## 7. Conclusion

We presented the description of an experimental installation for the test of the weak equivalence principle for neutrons. The device is a sensitive spectrometer of ultracold neutrons designed for precisely comparing the gain in kinetic energy of free falling neutrons with a quantum of energy $\hbar\Omega$ being transferred to the neutrons by a non-stationary device, a quantum modulator realized via a rotating phase grating. The UCN spectrometry is based on the use of multilayered interference filters of various types. They are applied for the primary UCN monochromatization, the extraction of the necessary line from a discrete spectrum obtained by diffraction of UCNs by a moving grating as well as for the suppressing of the background of very cold neutrons. For the analysis of the resulting neutron energies we used a new method. It is based on a periodical low frequency modulation of the neutron flux via a chopper and a sequential measurement of the count rate oscillation phase of a detector distant from the chopper-modulator by some path length. This method provides an accuracy in the measurement of the neutron velocity in the order of $\Delta V/V \approx 5\times10^{-4}$ during a measurement time of about 15 min.

The results of test experiments are presented and discussed. The primary problem of a insufficient luminosity was substantially solved and the collection rate of the statistical accuracy of the equivalence factor was improved to be t= $5\times10^{-3}$ per day. At the same time some systematic effects were found. Some of them may be easily excluded, but for the elimination of others more detailed investigations and analysis are necessary. Some possibilities to improve the device are shortly discussed.

Authors are very grateful to Tomas Brenner for outstanding technical support. This work was partly supported by the Russian Foundation for Basic Investigations (grants RFBI 09-02-000117a and 12-02-00902-a)

**References**

1. 1. A.W. Mc. Reinolds, Phys. Rev. **83** (1951) 172.
2. M. Kasevich and S. Chu. Phys. Rev. Lett. **67** (1991) 181
3. A. Peters, K. Y. Chung, and S. Chu. Nature **400** (1999) 849
4. A. Peters, K.Y. Chung and S. Chu. Metrologia **38** (2001) 25
5. S. Fray, C.A. Diez, Th.W. Hänsch, and M. Weitz, Phys. Rev. Lett. **93** (2004) 240404
6. J. B. Fixler, G.T. Foster, J.M. McGuirk, M.A. Kasevich. Science. **315** (2007) no 5808 74
7. J.W. Dabbs, J.A. Harvey, D. Paya, H. Horstmann, Phys. Rev. **139** B (1965) 757
8. H. Maier-Leibnitz, Z. Angew. Physik **14** (1962) 738
9. L .Koester, Z. Phys. **182** (1965) 328
10. L. Koester, Phys. Rev. D **14** (1976) 907
11. V.F. Sears, Phys. Rev. D **25** (1982) 2023
12. J. Schmiedmayer, Nuc. Instr. Meth. A **284** (1989) 59

13. L. Koester, W. Waschkowski, L.V. Mitsina et al. Phys. Rev. C **51** (1995) 3363.
14. L.V. Mitsyna, V.G. Nikolenko, S.S. Parzhitski, A.B. Popov, G.S. Samosvat, Eur. Phys. J. C (2005) 02134
15. R. Colella, A.W. Overhauser, S.A. Werner, Phys. Rev. Lett. **34** (1975)1472
16. J.-L. Staudenmann, S.A. Werner, R. Colella, A.W. Overhauser, Phys. Rev. A **21**(1980) 1419
17. K.C. Littrell, B.E. Allman, S.A. Werner, Phys.Rev. A **56** (1997)1767
18. G.Z. Adunas, E. Rodriguez-Milla, and D.V. Ahluwalia, arXiv:gr-qc/0006022 (2000) v1.
19. G. van der Zouw, M. Weber, J. Felber, et al. Nuclear Instr. Meth. A, **440** (2000) 568
20. V.-O. de Haan, J. Plomp, Ad A. van Well et al. Phys. Rev. A **89** (2014) 063611
21. V.V. Nesvizhevsky, H.G. Börner, A.K. Petukhov, et al. Nature **415** (2002) 297
22. V.V. Nesvizhevsky, H.G. Börner, A.M. Gagarski, A. K. Petukhov, et al., Phys.Rev. **67** (2003) 102002.
23. V.I. Luschikov and A.I. Frank. JETP Letters **28** (1978) 560
24. A. I. Frank, V. G. Nosov, JETP Letters **79** (2004) 313
25. T. Jenke, P. Geltenbort, H. Lemmel and H. Abele. Nature Physics **7** (2011) 468
26. E.O. Vezhlev, V.V. Voronin, I.A. Kuznetsov et.al., Phys. Part. and Nuc. Letters **10 (**2013) 357.
27. A.I. Frank, P. Geltenbort, M. Jentschel,et al. JETP Letters **86** (2007) 225
28. A.I.Frank, V.G.Nosov, Phys. Lett. A **188** (1994) 120
29. A.I. Frank, S.N. Balashov, I.V. Bondarenko, et al. Phys. Lett., A **311** (2003) 6
30. S.N. Balashov, I.V. Bondarenko, A.I. Frank, et al., Physica B **350** (2004) 246
31. A.I. Frank, P. Geltenbort, G. V. Kulin, et al. JETP Letters **81** (2005) 427
32. A. I. Frank, P. Geltenbort, G. V. Kulin, et al., JINR Communication P3-2004-207, Dubna, 2004 (in Russian).
33. Yu.N. Pokotilovskii, M.I. Novopoltsev, JINR communication, P3-81-821, 1981 (in Russian); M.I. Novopoltsev, Yu.N. Pokotilovskii, I.G. Shelkova, Nucl. Instrum. Methods. A. **264** (1988) 518
34. I.V. Bondarenko, V.I. Bodnarchuk , S.N. Balashov, et al., Phys. At. Nuc. **62** (1999) 721
35. A.I. Frank, S. V. Balashov, V.I. Bodnarchuk, Proc. SPIE **3767** (1999) 360
36. I.V. Bondarenko, A.I. Frank, S.N. Balashov, et al., NIM A **440** (2000) 591
37. A.I. Frank, V.I. Bodnarchuk, P. Geltenbort, et al., Phys. At. Nuc. **66** (2003) 1831
38. A.I. Frank, P. Geltenbort, M. Jentschel, et al. Phys. At. Nuc., **71** (2008) 1656
39. A. I. Frank, P. Geltenbort, M. Jentschel, et al., JETP Letters **93** (2011) 361
40. M. Maaza and D. Hamidi, Physics Reports **514** (2012) 177
41. A.A.Seregin, JETP **73** (1977) 1634
42. K.-A. Steinhauser, A. Steyerl, H. Schechenkofer and S.S. Malik, Phys. Rev. Lett. **44** (1980) 1306
43. A. Steyerl, W. Drexel, S.S. Malik, E. Gutsmiedle, Physica B **151**,36 (1988)
44. A.I. Frank, P. Geltenbort, M. Jentschel et al. Nucl. Instrum. Methods **A 611** (2009) 314
45. A. Steyerl, H. Nagel, F. Schriber, et al., Phys. Lett. A**116** (1986) 347.
46. A.I.Frank, P.Geltenbort, M.Jentschel, et al. International Seminar on Interaction of Neutrons with Nuclei (ISINN-19). Dubna, May 24-28, 2010. *JINR E3-2012-30*, Dubna, 98 (2012)
47. A.I. Frank, P. Geltenbort, M. Jentschel, et al., International Seminar on Interaction of Neutrons with Nuclei (ISINN-20). Alushta, Ukraine**,** May 21 – 26, 2012. *JINR E3-2013-22,* Dubna, 18 (2013)
48. A.I. Frank, P. Geltenbort, S.V. Goryunov, et al. International Seminar on Interaction of Neutrons with Nuclei (ISINN-21). Alushta, Ukraine**,** May 20 – 25, 2013. *JINR E3-2014-13,* Dubna, 67 (2014).